# Evolutionary Optimisation of Real-Time Systems and Networks

Leandro Soares Indrusiak, Robert I. Davis, Piotr Dziurzanski
Real-Time Systems Group, Department of Computer Science, University of York, United Kingdom

[leandro.indrusiak | rob.davis | piotr.dziurzanski] @york.ac.uk

*Abstract*—The design space of networked embedded systems is very large, posing challenges to the optimisation of such platforms when it comes to support applications with real-time guarantees. Recent research has shown that a number of inter-related optimisation problems have a critical influence over the schedulability of a system, i.e. whether all its application components can execute and communicate by their respective deadlines. Examples of such optimization problems include task allocation and scheduling, communication routing and arbitration, memory allocation, and voltage and frequency scaling. In this paper, we advocate the use of evolutionary approaches to address such optimization problems, aiming to evolve individuals of increased fitness over multiple generations of potential solutions. We refer to plentiful evidence that existing real-time schedulability tests can be used effectively to guide evolutionary optimisation, either by themselves or in combination with other metrics such as energy dissipation or hardware overheads. We then push that concept one step further and consider the possibility of using evolutionary techniques to evolve the schedulability tests themselves, aiming to support the verification and optimisation of systems which are too complex for state-of-the-art (manual) derivation of schedulability tests.

## I. INTRODUCTION

Applications such as autonomous vehicles, 5G mobile communications, live video processing and tele-surgery all require a heavy computational load to be performed within a bounded amount of time. Such applications are already deployed today, and are likely to become more pervasive in the near future. One way to enable their cost-effective deployment is via the use of real-time networked embedded systems, i.e. platforms based on multiple networked processors that are able to provide real-time guarantees to the application components running on and communicating over them.

In order to provide such real-time guarantees, system designers must configure the system in such a way that bounds the execution and communication time of application components. There are many approaches to that problem, ranging from architectural features at the processor and interconnect level, all the way to the allocation and scheduling of software tasks by the operating system or virtual machines. In this paper, we review a number of system-level configurations that have been shown to have a critical impact on the ability of the system to provide real-time guarantees.

Some typical system configurations affecting real-time behaviour are:

- task allocation and scheduling: defines when, and on which processor a task (or a part of a task) will be executed;
- communication routing, encoding and arbitration: defines when, in which form, and through which interconnect element(s) a task will communicate with other tasks;
- memory allocation: defines which areas of system memory are allocated for a given task to store its binaries, data and communication buffers;
- voltage and frequency scaling: defines which clock frequency (and the respective level of voltage) will be assigned to each processor and interconnect element at each point in time.

Given the complexity and sheer size of the design spaces characterised by such configurations, it is unlikely that a solution can be found which is simultaneously optimal for all problems. Actually, for reasonably complex systems, it is already infeasible to find optimal solutions to a single one of those optimization problems, let alone to all of them. Most research approaches use heuristics to navigate over the design spaces in search for optimised solutions. Evolutionary algorithms are a particularly suitable heuristic for this kind of optimisation problem, mainly due to their potential parallelism and ability to navigate design spaces without making assumptions about their respective solution spaces.

## II. EVOLUTIONARY OPTIMISATION OF REAL-TIME SYSTEMS

Evolutionary optimisation is the application of an evolutionary algorithm to iteratively uncover improved solutions to an optimisation problem. Such an algorithm is heuristic in nature, meaning that there is no guarantee that it will ever find an optimal solution, or that it will identify a solution as optimal if it is found. Nevertheless, such approach is widely used in a variety of optimisation problems in science and engineering [1][2], where a sufficiently good solution is acceptable despite of being suboptimal.

To solve such optimisation problems, evolutionary algorithms represent potential solutions as a chromosome, which is effectively a vector containing the values adopted by that particular solution for all the optimisation variables of the problem at hand. For example, if the problem is a task mapping problem, the chromosome may comprise a vector of the allocations of each of the tasks, e.g. a reference to the processor that will execute each of them (as shown in Figure 1.a for a set of tasks $\tau_1$-$\tau_n$ and a set of processors $\pi_a$-$\pi_z$). The algorithm operates over a population of individuals, each represented by one such chromosome. The initial population can be randomly generated, but subsequent generations are produced by applying crossover and mutation operations over the fittest chromosomes of the preceding one (Figure 1.b). This requires the definition of appropriate crossover and mutation operations (which must produce valid chromosomes out of existing ones), a fitness function (to rank the chromosomes) and a selection function (which chooses among the best-ranked chromosomes which ones will be used as the basis for the next generation). After a number of generations, the population will contain individuals of improved fitness and, ideally, at least one of them which fulfils the fitness requirements of the optimisation problem.

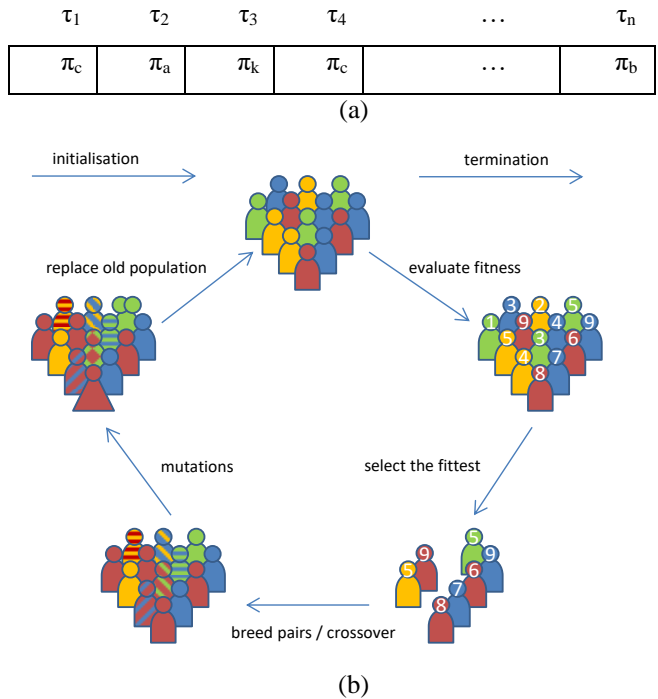

Fig. 1. Evolutionary optimisation: (a) chromosome format and (b) evolutionary algorithm.

Different optimisation problems will obviously require different fitness functions, and most will be sensitive to different chromosome formats, population sizes, mutation and crossover styles and rates. For example, a detailed study on how such aspects affect the convergence of evolutionary algorithms in real-time systems can be found in [17].

Over the past decades, evolutionary algorithms have been widely used to solve task scheduling in real-time systems [3], [4], [5]. As it is the case in many classic real-time systems research, inter-task communications are not considered (or implicitly considered as part of computation time). In this paper, we address networked embedded systems, so both computation and communication resources and their respective load (task, network packets) are explicitly considered. Therefore, we don't review evolutionary optimisation of classic computation-only models, and instead assume them to be special cases of the systems we address next.

When it comes to networked embedded platforms, Ascia et al. [6] were among the first to use evolutionary optimisation, but they did not address real-time performance guarantees. They proposed an evolutionary algorithm to minimise energy and average performance of multiprocessor platforms based on Networks-on-Chip. Their approach used system simulation as the optimisation fitness function, i.e. as the way to accurately obtain average performance figures for each configuration, and thus guide the evolutionary search towards optimised results. Their extensive set of experiments provided important insights on the potential of such techniques, but also highlighted the heavy cost of performing thousands of simulations, which could take several hours or even days.

Mesidis and Indrusiak [7] presented the first approach to couple evolutionary algorithms and real-time schedulability models for networked embedded systems. They proposed the use of schedulability tests as a fitness function, aiming to find configurations that fulfil hard real-time constraints (i.e. tasks and communications never miss their deadlines). Besides showing the successful application of the technique, they have also shown that schedulability models are more suitable than simulation as fitness functions for evolutionary optimisation, as they perform orders of magnitude faster and can therefore be applied to thousands of individuals across hundreds of generations. Following this trend, Ayari et al. [8] proposed the use of custom genetic operators to make better use of schedulability models as search guides within an evolutionary optimisation pipeline.

The possibilities of multi-objective evolutionary optimisation were investigated by Sayuti and Indrusiak [9], combining a schedulability model [10] and an energy macro-model [11]. The heuristic was able to evolve configurations that fulfil hard real-time constraints and simultaneously minimise energy dissipation by the platform.

All the approaches above address a single type of optimisation, namely task allocation. Figure 2 depicts the typical evolutionary pipeline [7], which has been extended in the subsequent works. Chromosomes encode task allocations as shown in Figure 1.a. Again, the initial population is randomly generated, and is then diversified by a pipeline of genetic operations (selection, crossover and mutation), producing an offspring population. The offspring population is evaluated according to a fitness function based on a schedulability metric (e.g. the number of fully-schedulable tasks and communications in each alternative allocation [7]). The fittest chromosomes then become a part of (or replace) the parent population for the next iteration, until some termination criteria is met.

Further research has incorporated other types of configurations into the task allocation problem described above. Sayuti et al. [12] have extended the evolutionary pipeline above to simultaneously optimise task allocation and communication encoding. The aim being to encode data exchanged by tasks allocated to processors that are not in the network neighbourhood of each other, so that the transition activity on the interconnect wires is minimised, and so, as a consequence, is the energy dissipated. Similarly, Sayuti and Indrusiak [13] have simultaneously optimised task allocation and communication arbitration by encoding the communication priority together with the allocation information in each chromosome.

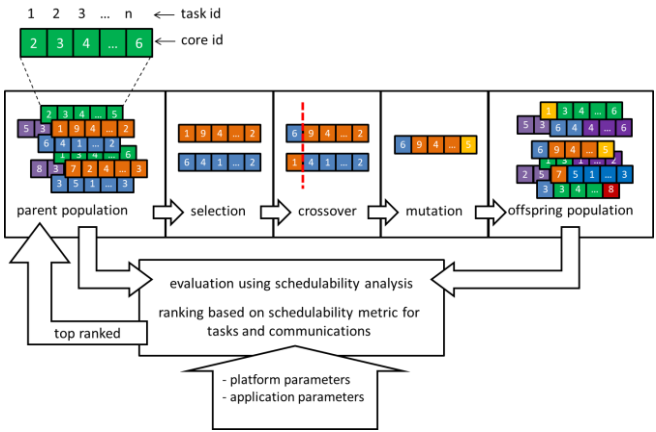

Fig. 2. Evolutionary approach for task allocation

Additional changes to the chromosome representation allowed Still and Indrusiak [14] to simultaneously optimise task allocation and memory allocation. With multi-objective fitness functions, they could separately optimise the allocation of task binaries, local data and context, and their communication buffers. Also by changing chromosome representation, Indrusiak et al. [15] have simultaneously optimised task allocation and communication routing, aiming to increase resilience to side-channel attacks over the on-chip interconnect.

A more sophisticated approach, which required changes to the fitness function rather than to the chromosome representation, was proposed by Sayuti and Indrusiak [16]. Instead of defining the fitness of a configuration to be the number of fully-schedulable tasks and communications in each alternative allocation (as proposed in [7]), this work used the lowest multiplicative factor to the nominal frequency of the processors and interconnect network that would render the system fully schedulable. This factor is referred to as the *breakdown frequency* of a particular configuration. The breakdown frequency of a poor configuration of the system would likely be very high (i.e. much larger than 1, which corresponds to the nominal frequency), because unless the system was significantly overclocked it would not be fully schedulable. On the other hand, a good configuration of the system might render it fully schedulable at the nominal frequency or lower (i.e. $0 <$ breakdown frequency $< 1$). Despite being more costly, as it required a binary search over the set of supported frequency settings, this approach allowed a simultaneous optimisation of task allocation and frequency scaling, effectively dominating the prior approaches reported in [7] and [9].

Also by changing the fitness function, Dziurzanski et al. [17] optimised task allocations in multi-mode applications (i.e. applications where the computation and communication loads can change over its lifetime). Their approach evolved task allocations for each of the application modes in such a way that they are fully schedulable, and that the overhead of changing modes is minimised (i.e. the number of migrated tasks and the amount of data and context information to be migrated).

## III. Evolving Schedulability Tests

As discussed in the previous section, all evolutionary optimisation approaches supporting real-time guarantees use some sort of schedulability test as the fitness function guiding the search process. Such a test is applied to each and every chromosome in the population, and determines the fitness of each of them, e.g. as the percentage of tasks and network packets that are fully schedulable under the respective system configuration. It is therefore difficult to argue against the statement that an optimisation process is only as good as its underlying schedulability test. This limits the potential application of evolutionary optimisation to real-time systems and networks with known schedulability tests, and tests that are sufficiently tight (i.e. sufficient tests that are not overly restrictive).

For simple real-time systems and networks, there are mature and well-established schedulability tests which were handcrafted over the years by the research community. Most of those tests are supported by informal proofs, which are also devised by manually putting together logical arguments so that members of the community can check and be convinced of their correctness. More recently, initiatives addressing the use of proof mechanisation and automatic proof checkers have been proposed [18], but so far they are only able to automate and verify proofs for schedulability tests of relatively simple systems.

As the complexity of real-time systems and networks increases, it is likely that manual construction and verification of schedulability tests will not always produce correct and useful tests. There is already evidence to support such a statement, as several schedulability tests accepted and used by the community (some of them for several years) were recently shown to be flawed [19], [20], [21]. The use of proof mechanisation is likely to improve that picture, but it may take many years of research before it reaches the level of complexity of state-of-the-art networked real-time systems.

We therefore advocate the use of evolutionary mechanisms to support the creation and verification of schedulability tests. We follow a similar pipeline as the one described in Figure 1.b, but where each individual is a schedulability test. Our current research is addressing ways to represent schedulability tests as chromosomes, devising mutations and crossover operators that operate over those chromosomes while producing semantically valid offspring tests, and investigating fitness functions that can rank different tests within a population from the point of view of correctness and tightness. Additional details about our approach are beyond the scope of this paper, and most of it is still work in progress, but the intuition behind the approach is shown in Figure 3.

In Figure 3.a, we represent the set of all possible configurations of a real-time system or network as a light blue rectangle. Within that rectangle, we represent the subset of configurations that are fully schedulable as a dark blue rectangle. We then represent schedulability tests as rounded dashed rectangles: a necessary schedulability test (which includes all schedulable configurations, but also some unschedulable ones), a sufficient schedulability test (which includes only schedulable configurations, but not all of them) and an exact schedulability test (which follows exactly the boundary of our dark blue rectangle). The scenario we are interested in, however, addresses real-time systems for which such tests still do not exist, and the exact boundaries of the dark blue rectangle are unknown.

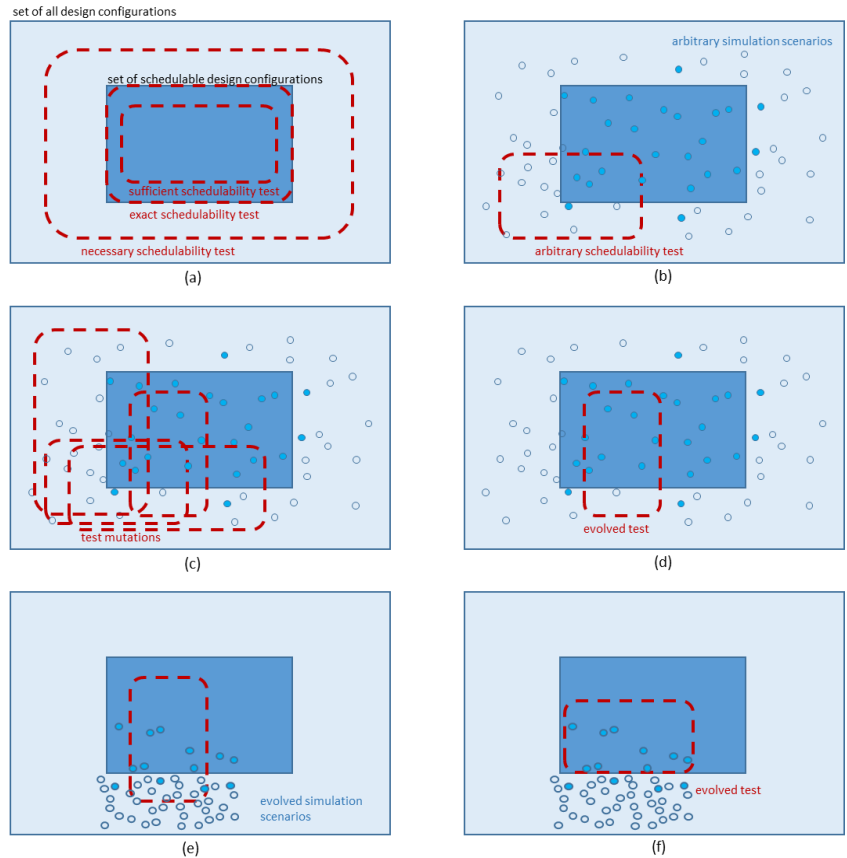

Fig. 3. Evolution of schedulability tests

The approach we advocate starts with a population of arbitrary schedulability tests. As explained in Section II, such an initial population could be produced randomly, and therefore will be composed of arbitrary tests such as the one shown in Figure 3.b. Such initial tests will probably not be very useful in discriminating schedulable and unschedulable configurations of a system, but some of them will be more useful than others. With the help of an arbitrary set of simulation scenarios (shown in Figure 3 as small circles), it is possible to associate a fitness to a schedulability test. A single simulation scenario cannot prove that a particular configuration is schedulable, but it can show that it is unschedulable simply by detecting that at least one task or network packet has missed its deadline. We therefore use such simulation scenarios to show how well (or how poorly) a given schedulability test performs. In the example shown in Figure 3.b, one can infer that the fitness of the depicted test is not particularly good, since it marks as schedulable (i.e. within its bounds) a number of scenarios that simulation has shown as unschedulable (i.e. hollow circles).

Following the evolutionary approach, we then create a number of offspring schedulability tests by applying mutation and crossover operators. This is shown in Figure 3.c, where a number of mutations of the initial test are depicted. Using the fitness metric mentioned above, it is possible to see that the fittest among of those mutations is the one that surrounds the least number of unschedulable configurations uncovered by simulation. The fittest tests are then passed on to the next generation (Figure 3.d), and the whole evolutionary process iteratively applied until an acceptable schedulability test can be found.

To reduce the computational cost of running simulation scenarios, it is also possible to evolve the set of scenarios together with the evolution of the schedulability tests. That way, each population of tests will have their fitness evaluated by scenarios that were evolved to uncover "counterexamples" (configurations wrongly deemed as schedulable by the tests), as shown in Figure 3.e. Through careful inter-play between the evolution of schedulability tests and the sets of simulation scenarios, we aim to uncover tests that are sufficient (i.e. completely within the dark blue rectangle) but not overly restrictive (i.e. not much smaller than the dark blue rectangle), as shown in Figure 3.f.

## IV. PRELIMINARY RESULTS

This section presents a proof-of-concept case study on the use of evolutionary algorithms to synthesise response time formulae for the schedulability analysis of messages on Controller Area Network (CAN). The popularity of CAN and the hard requirement of its message-set schedulability resulted in the derivation of several proven exact tests and sufficient tests of pseudo-polynomial complexity [19]. These tests determine $R_i$, the worst-case response time of message $i$ by analysing the influence of all instances of higher priority messages in the same message-set following the worst-case message release scenario. A pessimistic test produces values of $R_i$ that are higher than the actual response time of a message, which is still useful because it provides a sound upper bound. However, an optimistic test produces values of $R_i$ that are lower than the actual response time, which is unsound in worst-case analysis and should be avoided at all costs.

Each message $i$ is characterised by its worst-case execution time $C_i$, period $T_i$, jitter $J_i$ and deadline $D_i$. Since immediately before the initial release of message $i$, the longest message of a lower priority can begin transmission, the longest transmission time of any message of lower priority, denoted as $B_i$, also needs to be considered. Equation (1) presents a popular sufficient test **S1** given in [19] reformulated as a single expression.

$$R_i^{m+1} = J_i + C_i + max(B_i, C_i) + \sum_{k \in hp(i)} \left( \left\lfloor \frac{R_i^m - J_i - C_i + J_k}{T_k} \right\rfloor + 1 \right) C_k \quad (1)$$

where $hp(i)$ denotes the set of messages with higher priorities than message $i$. Equation (1) has pseudo-polynomial time complexity as the final solution needs to be found using fixed-point iteration. Since $R_i$ is upper-bounded by $D_i$, equation (1) can be replaced with its closed-form equivalent that provides a more pessimistic approximation, as presented in the following equation.

$$R_i = J_i + C_i + max(B_i, C_i) + \sum_{k \in hp(i)} \left( \left\lfloor \frac{D_i - J_i - C_i + J_k}{T_k} \right\rfloor + 1 \right) C_k \quad (2)$$

Equation (2) can be further simplified by removing the subtraction of jitter and worst-case execution time in the numerator, leading to a further more pessimistic approximation given by the following equation.

$$R_i = J_i + C_i + max(B_i, C_i) + \sum_{k \in hp(i)} \left( \left\lfloor \frac{D_i + J_k}{T_k} \right\rfloor + 1 \right) C_k \quad (3)$$

Equations (1)-(3) will be used as a baseline to demonstrate the potential of the envisioned evolutionary approach to derive schedulability tests.

To evaluate the possibility of evolving response time formulae for the CAN schedulability problem, we decided to use EpochX, an open source genetic programming framework[2], because of its basic functionality supporting grammatical evolution. We created a BNF grammar description including variables and operators that typically appear in response time formulae, and allowed EpochX to evolve parse trees, which in turn generate syntactically valid equations based on that grammar. The fitness of each expression was evaluated based on how well it calculates the response time of 2600 CAN messages divided into 135 different message-sets. In this case study, the smaller the fitness, the better, meaning that results produced by the evolved equation are close to the actual response time values. (To avoid optimistic tests, expressions that produce results that are lower than the real response times are heavily penalised by the fitness function).

The structure of the formulae generated in this case study resemble, in general, the structure of the baseline equations. For example, one of the formulae with the best fitness is of the following form.

---

[2] https://www.epochx.org/

$$R_i = J_i + C_i + B_i + \sum_{k \in hp(i)} \left( \left\lceil \frac{min(max(min(max(J_i, C_i), J_k),}{T_k} \right. \right.$$
$$\left. \left. \frac{max(J_k, min(C_i, C_k) - C_i)) + R_i, T_i)}{} \right\rceil \right) C_k$$
$$. \quad (4)$$

Some parts of equation (4) have a good resemblance to the known sufficient tests, e.g., equation (1). Despite missing some elements, additional terms seem to compensate their omission in the final evaluation. Note that the generated formulas can be simplified by using well-known relations between message parameters, for example $T_k \geq D_k \geq C_k$. Defining and automatically applying a full set of simplifying rules is left as future work.

Figure 4 shows the (normalised) fitness comparison between the baseline formulae (1)-(3) and the best equation generated by the proposed evolutionary approach (4). The generated formula leads to the fitness value which is about 8 times worse than the least pessimistic formula (1), but still almost the same fitness as schedulability test (2) and 30% better than schedulability test (3). The generated formula did not produce optimistic response times for any of the 2600 messages used in the comparative assessment. The generated formula can thus be regarded as having a similar accuracy to the formulae generated by human ingenuity and could with further checking potentially be suitable for use in practice. Of course, that is not the case for this particular scenario as exact schedulability tests are known for CAN, and our goal here was simply to highlight the efficacy of the proposed evolutionary approach on this problem.

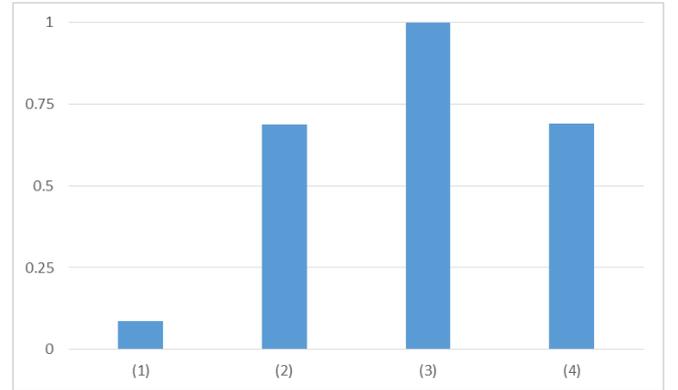

Fig. 4. Normalised fitness values of baseline (1)-(3) and generated (4) formulae obtained on an example message set (lower is better)

## V. CONCLUSIONS

Evolutionary optimisation has been successfully applied to real-time systems and networks, achieving improvements to multiple aspects of their operation (e.g. energy dissipation, memory/buffer overheads) while at the same time guaranteeing full schedulability. To the best of our knowledge, in all published cases, the evolutionary process is guided by existing schedulability tests.

The creation and validation of schedulability tests for complex real-time systems and networks is increasingly difficult, so we speculate about the use of an evolutionary pipeline to support that process. If successful, such an approach will certainly contribute towards schedulability analysis for the next generation of real-time systems and networks.

The proof-of-concept described here encourages us to pursue this path. However, the case study attacked a problem with known schedulability equations, and used known worst-case response times as the basis for its fitness function. The problem will be much harder when the fitness function has to rely on high watermark or approximate values instead, which will be the case for every problem with as yet unknown schedulability tests.

ACKNOWLEDGEMENTS

The authors are thankful for the contributions of Jerry Swan, for suggestions of techniques and tools used in the reported experimental work. The authors also acknowledge the support of the EU H2020 SAFIRE project (Ref. 723634).